\documentclass[useAMS,usenatbib]{mn2e}
\usepackage{graphicx}

\newcommand{\apj}{ApJ}

\newcommand{\aap}{A\&A}
\newcommand{\aaps}{A\&AS}
\newcommand{\aj}{AJ}
\newcommand{\mnras}{MNRAS}

\newcommand{\pasp}{PASP}
\newcommand{\apjs}{ApJS}
\newcommand{\MC}{\multicolumn}

\newcommand{\NGC}{NGC\,6822}
\newcommand{\Sym}{NGC\,6822~\mbox{SySt--1}}
\newcommand{\ICSym}{IC\,10~\mbox{SySt--1}}
\DeclareRobustCommand{\ion}[2]{%
\relax\ifmmode
\ifx\testbx\f
{\mathrm{#1\,\textsc{#2}}}\else
{\mathrm{#1\,\mathsc{#2}}}\fi
\else\textup{#1\,{\mdseries\textsc{#2}}}%
\fi}

\title[Discovery of the first symbiotic star in \NGC]
{Discovery of the first symbiotic star in \NGC\footnotemark[0]\thanks{
Based on observations made with the Southern African Large Telescope (SALT)
and the Infrared Survey Facility (IRSF).}
}
\author[A. Y. Kniazev et al.]{%
Alexei Y.\ Kniazev,$^{1,2}$\thanks{E-mail: akniazev@saao.ac.za (AYK)}
Petri V\"ais\"anen,$^{1,2}$
Patricia A. Whitelock,$^{1,3}$
\newauthor
John W. Menzies,$^{1}$
Michael W. Feast,$^{1,3}$
Eva K.\ Grebel,$^{4}$
David A.H.\ Buckley,$^{1,2}$
\newauthor
Yas Hashimoto,$^{1,2}$
Nicola Loaring,$^{1,2}$
Encarni Romero-Colmenero,$^{1,2}$
\newauthor
Ramotholo Sefako,$^{1,2}$
Eric B. Burgh,$^{5}$
Kenneth Nordsieck$^{5}$\\
\rule{-4pt}{20pt}
$^{1}$South African Astronomical Observatory, PO Box 9, 7935 Observatory, Cape Town,
South Africa.\\
$^{2}$Southern African Large Telescope Foundation, PO Box 9, 7935 Observatory, Cape Town,
South Africa.\\
$^{3}$Astronomy Department, University of Cape Town, 7701 Rondebosch,
South Africa.\\
$^{4}$Astronomisches Rechen-Institut, Zentrum f\"ur Astronomie Heidelberg,
M\"onchhofstr.\ 12--14, D-69120 Heidelberg, Germany.\\
$^{5}$Space Astronomy Laboratory, University of Wisconsin, Madison, WI 53706,
USA. \\
}

\begin{document}

\date{Accepted 2009 ??. Received 2009 ??; in original form 2009}

\pagerange{\pageref{firstpage}--\pageref{lastpage}} \pubyear{2009}

\maketitle

\label{firstpage}

\begin{abstract}
We report the discovery of the first symbiotic star ($V=21.6$, $K_S=15.8$
mag) in the Local Group dwarf irregular galaxy \NGC.
This star was identified during a spectral survey of H$\alpha$ emission-line
objects using the Southern African Large Telescope (SALT) during its
performance-verification phase. The observed strong emission lines of
\ion{H}{i} and \ion{He}{ii} suggest a high electron density and
$T^* < 130\,000$~K for the hot companion. The infrared colours
allow us to classify this object as an S-type
symbiotic star, comprising a red giant losing mass to a compact
companion. The red giant is an AGB carbon star, and a
semi-regular variable, pulsating in the first overtone with a period of 142
days. Its bolometric magnitude is $M_{bol}=-4.4$ mag.

We review what is known about the luminosities of extragalactic symbiotic
stars, showing that most, possibly all, contain AGB stars. We suggest that a
much larger fraction of Galactic symbiotic stars may contain AGB stars than
was previously realised.
\end{abstract}


\section{Introduction}

The galaxies of the Local Group (LG) and its immediate surroundings are of
particular interest.  Their proximity permits us to study many aspects
of galaxy evolution in great detail, and this detailed information is in
turn used to constrain studies of high-redshift galaxy formation.
In particular, individual \ion{H}{ii} regions, supergiants, planetary nebulae
and emission-line stars such as symbiotic stars in the LG are available
for high S/N spectrophotometry with present-day
large telescopes; this information can be used  
to derive accurate chemical abundances.  The goal is to use these
abundances, together with colour-magnitude studies of stellar populations in 
these galaxies, to unveil the detailed chemical
evolution and star formation history of the LG galaxies and to answer 
questions such as whether dwarf spheroidals and ellipticals are the end result 
of gas-rich dwarf irregulars \citep[e.g.,][]{GGH03}.
At the same time we should be able to obtain a
more detailed understanding of stellar evolution in different environments. 

We have an on-going project using several different telescopes to determine
accurate chemical abundances of PNe and \ion{H}{ii} regions in a sample of the nearest
dwarf galaxies \citep{Sextans, Ket05, Fornax, IC10, Kni08} that we have
expanded to study emission-line stars. Here we report
the spectroscopic discovery of the first symbiotic star
in the well studied LG dwarf galaxy, \NGC, which is both gas-rich
and exhibits continuing star formation
\citep[see e.g.,][ and references therein]{Lee06}. We will call this star
\Sym\ following the naming pattern that was suggested by
\citet{IC10_SySt} for the first symbiotic star found in the LG galaxy IC\,10.

\begin{figure}
    \begin{center}
    \includegraphics[angle=-0,width=9.0cm]{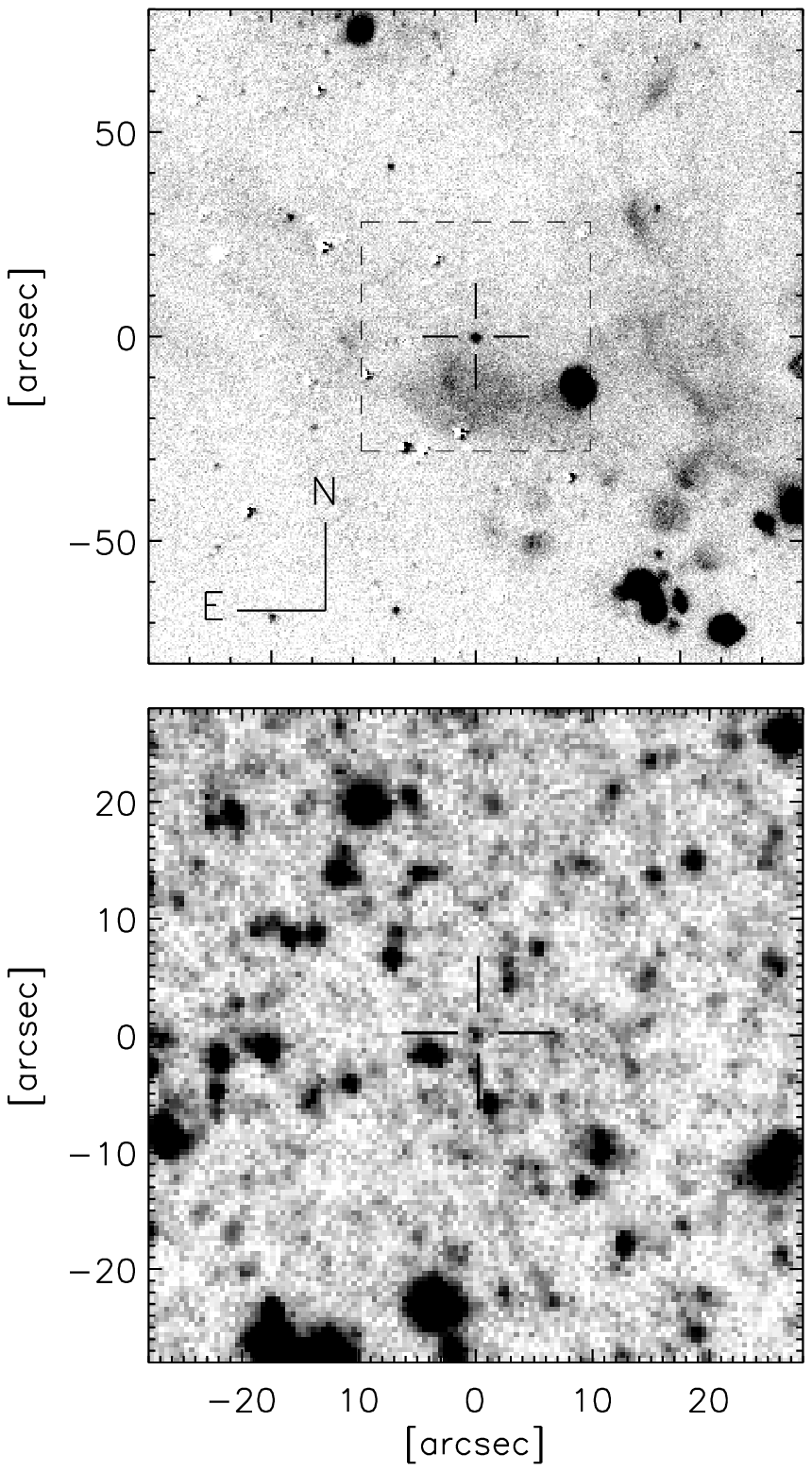}
    \caption{
    Continuum subtracted H$\alpha$ (top) and $V$-band (bottom) images
    \citep{Ken03} of the area of interest in \NGC.
    The dashed line in the top panel outlines the area illustrated in
    the lower panel. The
   Symbiotic star \Sym\ (\mbox{$\alpha_{2000.0}$ = 19:44:53.88},
   $\delta_{2000.0}$ = $-$14:51:46.8)
    is located at the centre of each image.
    For the H$\alpha$ image the contrast is adjusted to highlight both 
    bright and faint H$\alpha$ emission.
    Black objects on the image indicate bright sources.
    \label{fig:FC}}
    \end{center}
\end{figure}

Symbiotic stars are interacting binaries, in which an evolved giant
transfers material to a much hotter, compact companion, usually a white dwarf,
more rarely a neutron star or a main-sequence star. Mass-transfer is often
via an accretion disk.
High excitation emission lines are produced in the surrounding nebulosity,
excited by the hot source. Symbiotics are of interest among other things as
possible progenitors of type Ia supernovae \citep[e.g.,][]{MR92}.

We have adopted a distance modulus of $(m-M)_0 = 23.31$ mag for NGC\,6822
which is based on classical Cepheids \citep{Gier06}. There is some remaining
uncertainty in this value resulting from uncertainty in the adopted LMC
distance (Gieren et al. assume $(m-M)_0 = 18.5$ mag), reddening correction
uncertainties, and any corrections that might be necessary for the
difference in metallicities between the LMC and NGC\,6822. However,
uncertainties of the order of 0.1 or 0.2 mag will not affect any of our
conclusions regarding the nature of the stars discussed.

\section{Observations and Data Reduction}
\label{txt:Obs_and_Red}

%
%
\begin{table}
\centering{
\caption{Line intensities of \Sym}
\label{t:Intens}
\begin{tabular}{lcc} \hline
\rule{0pt}{10pt}
$\lambda_{0}$(\AA) Ion                  & F($\lambda$)/F(H$\beta$)& EW($\lambda$) \\ \hline
4101\ H$\delta$\                        & 0.25$\pm$0.07           & ~45$\pm$11    \\
4340\ H$\gamma$\                        & 0.40$\pm$0.05           & ~42$\pm$6~    \\
4686\ He\ {\sc ii}\                     & 0.36$\pm$0.05           & ~60$\pm$8~    \\
4861\ H$\beta$\                         & 1.00$\pm$0.09           & 210$\pm$18    \\
5876\ He\ {\sc i}\                      & 0.13$\pm$0.04           & ~11$\pm$4~    \\
6563\ H$\alpha$\                        & 7.52$\pm$0.55           & 753$\pm$50    \\
6678\ He\ {\sc i}\                      & 0.05$\pm$0.04           & ~~4$\pm$3~    \\
6830\ Raman                             & 0.23$\pm$0.06           & ~12$\pm$4~    \\
  & & \\
F(H$\beta$)$^a$\          & \MC {2}{c}{7.6$\pm$0.4}   \\
\hline
\MC{3}{l}{$^a$ in units of 10$^{-16}$ ergs\ s$^{-1}$cm$^{-2}$.}\\
\end{tabular}
 }
\end{table}

\subsection{Spectroscopy}

\Sym\ was identified as a strong emitter in a continuum subtracted
H$\alpha$ image from the SINGS 5th data release \citep{Ken03}, retrieved
from the NED database. The $V$ (also from SINGS) and H$\alpha$ finding
charts are presented in Fig.~\ref{fig:FC}. The field of view covers
80$\times$80 square arcsec and shows stars as well as compact and extended
H$\alpha$ emission around the target.

\begin{figure}
    \begin{center}
    \includegraphics[angle=-90,width=8.5cm,clip=]{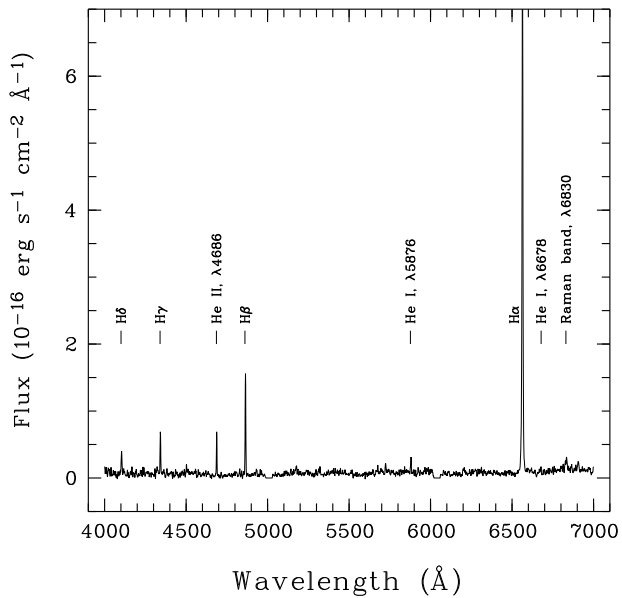}
    \caption{
    The observed spectrum of \Sym. All detected emission lines are shown.
    \label{fig:SySt}}
    \end{center}
\end{figure}

Spectral observations of selected H$\alpha$ emission sources in \NGC\ 
were taken at the prime focus of the newly available 10-m class
{\em Southern African Large Telescope} \citep[SALT;][]{Buck06,Dono06}
during its performance-verification stage while commissioning
the multi-object spectroscopic mode of the {\em Robert Stobie Spectrograph}
\citep[RSS;][]{Burgh03,Kobul03}.
Data for the field with \Sym\ were obtained on 2006 October 20. Two
exposures, of 1150~s and 960~s length respectively, fitting within a single
SALT visibility track, were taken in seeing conditions of approximately 1
arcsec. The RSS pixel scale is 0\farcs129 and the effective field of view
8\arcmin\ in diameter. We utilized a binning factor of 2, to give a final
spatial sampling of 0\farcs258 pixel$^{-1}$. The Volume Phase Holographic
(VPH) grating GR900 was used to cover the spectral range 4000--7000 \AA\
with a final reciprocal dispersion of $\sim$0.95 \AA\ pixel$^{-1}$ and a
spectral resolution FWHM of 5--6 \AA. Spectra of ThAr comparison arcs were
obtained to calibrate the wavelength scale.  The spectrophotometric standard
stars G~93-48 and Hiltner~600 were observed for relative flux calibration.
Data reduction was carried out in the standard manner described in
\citet[][]{Kni08}. Since SALT has a variable pupil size, an absolute
flux calibration is not possible using spectrophotometric standard stars. To
calibrate absolute fluxes we used objects visible in our target field and
the corresponding PN fluxes from \citet{Leisy05} and H$\alpha$ fluxes taken
from images of the `Survey of Local Group Galaxies Currently Forming Stars'
\citep{Mas06,Mas07}. The resulting reduced, extracted and flux calibrated
spectrum of \Sym\ is shown in Fig.~\ref{fig:SySt}. All emission lines in the
SALT spectrum were measured applying procedures described in detail in
\citet{SHOC}. All the detected emission lines, their equivalent widths and
relative strengths compared to the measured flux of the H$\beta$ emission
line, are presented in Table~\ref{t:Intens}.

\subsection{Infrared Imaging}

Fifteen observations of NGC\,6822 have been obtained over a period of almost
3.5 years as part of a survey of AGB stars in Local Group galaxies (Feast,
Whitelock, Menzies, in preparation). This programme uses the Japanese-South
African 1.4-m Infrared Survey Facility (IRSF) at Sutherland.
Simultaneous $J,H$ and $K_S$ images are obtained with the
SIRIUS camera, which has a field of view of about $7.5 \times 7.5$ square
arcminutes and a plate scale of 0.45 arcsec/pixel.
The data for the putative symbiotic and its close neighbour have been
extracted from the database. The $JHK_S$ photometry
was put on the 2MASS system with reference to about 70
stars in common in the field containing the symbiotic. No colour equations
were used. This photometry is listed in Table~\ref{t:phot}.

\begin{table}
\centering
\caption{Infrared photometry for \Sym.}
\label{t:phot}
\begin{tabular}{cccc}
\hline
JD--2450000 & $J$             & $H$            & $K_S$  \\
(day)       &(mag)            & (mag)          & (mag)  \\
\hline
 2353.49644& 17.261$\pm$0.028 &16.300$\pm$0.019& 15.750$\pm$0.025 \\
 2436.50394& 17.362$\pm$0.057 &16.402$\pm$0.039& 15.974$\pm$0.052 \\
 2441.50423& 17.529$\pm$0.043 &16.450$\pm$0.028& 15.892$\pm$0.039 \\
 2442.50430& 17.535$\pm$0.059 &16.540$\pm$0.041& 15.965$\pm$0.052 \\
 2507.34437& 17.064$\pm$0.030 &16.182$\pm$0.024& 15.675$\pm$0.030 \\
 2809.42608& 17.099$\pm$0.030 &16.237$\pm$0.025& 15.727$\pm$0.025 \\
 2529.29629& 17.154$\pm$0.025 &16.225$\pm$0.019& 15.719$\pm$0.021 \\
 2882.39648& 17.486$\pm$0.035 &16.443$\pm$0.025&       ...        \\
 3173.48911& 17.583$\pm$0.039 &16.621$\pm$0.024& 15.935$\pm$0.023 \\
 3243.39302& 17.349$\pm$0.027 &16.363$\pm$0.018& 15.770$\pm$0.017 \\
 3259.30017& 17.376$\pm$0.024 &16.351$\pm$0.017& 15.836$\pm$0.020 \\
 3260.30002& 17.366$\pm$0.021 &16.360$\pm$0.016& 15.840$\pm$0.019 \\
 3293.32443&       ...        &16.447$\pm$0.021& 15.947$\pm$0.041 \\
 3531.58939& 17.270$\pm$0.026 &16.320$\pm$0.019& 15.748$\pm$0.018 \\
 3533.44826& 17.336$\pm$0.027 &16.310$\pm$0.018& 15.779$\pm$0.019 \\
 3612.35759& 17.476$\pm$0.020 &16.464$\pm$0.014& 15.888$\pm$0.019 \\
\hline
\end{tabular}
\end{table}

\section{Results and Discussion}
\label{txt:disc}

\subsection{Astrometry and photometry}
\label{txt:ast_phot}

The images from \citet{Ken03} are astrometrically calibrated, and we 
derived the position \mbox{$\alpha_{2000.0}$ = 19:44:53.88} and
$\delta_{2000.0}$ = $-$14:51:46.8 with an error smaller than 1 arcsec
for \Sym.

From the calibrated SALT spectrum, using methods described in \citet{Sextans},
we calculated the $V$-band magnitude of this star, $V = 21.65\pm0.25$ mag,
which translates to $M_V\sim-2$ mag at our assumed distance modulus of 23.3 mag.
With the 4-m telescope of Cerro Tololo Inter-American
Observatory \citet{Mas07} obtained photometric data for the resolved stellar 
population of several dwarf galaxies with active star formation. We identify
\Sym\ as their emission-line star J194453.88-145146.6, which has $V=21.62$
mag ($(B-V)=1.20$ mag and $(U-B)=-0.18$ mag). Given that this is a symbiotic
star we would expect $V$ to vary; nevertheless our derived $V$ mag agrees
well with the published value.

\begin{figure}
\includegraphics[width=8.5cm]{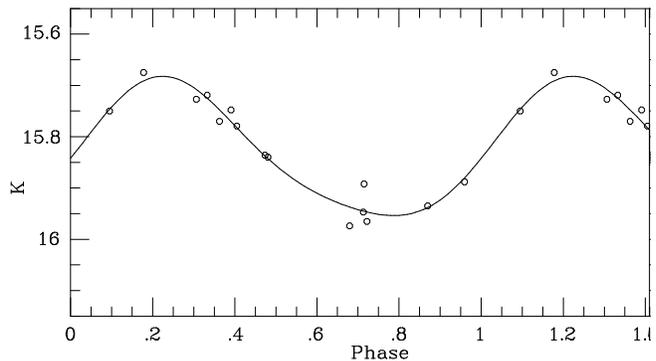}
 \caption{The $K$ light-curve for \Sym\ arbitrarily
	  phased on 142 days. The line shows the
	  best fitting 2nd order sine curve.}
 \label{fig_lc}
\end{figure}

A Fourier analysis of the 15 $H$ and 14 $J$ and $K_S$ observations yields a
clear period of $142\pm3$ days. The $K$ light-curve, folded on this period
is shown in Fig.~\ref{fig_lc} and the Fourier mean magnitudes and
peak-to-peak amplitudes are listed in the first two lines of
Table~\ref{t:ir}. The period and amplitudes suggest a semi-regular (SR)
variable and in order to make a comparison with published data we put our
data onto the SAAO system, following Carpenter's (2001) prescription as
given in the updated web
page\footnote{www.astro.caltech.edu/$\sim$jmc/2mass/v3/transformations/}.
The transformed magnitudes are given in the third line of Table~\ref{t:ir},
while the fourth lists the values after correction for extinction (see
section 3.2) according to the reddening law given by \citet{Glass99}.

\begin{table}
\centering
\caption{Mean Infrared photometry for stars of interest.}
\label{t:ir}
\begin{tabular}{cccl} 
\hline
$J$   & $H$   & $K_S/K$ & Description\\
(mag) &(mag)  & (mag)   &            \\
\hline
\multicolumn{4}{l}{\Sym}\\
\hline
17.34 & 16.37 & 15.82   & Fourier mean\\
0.48  & 0.35  & 0.27    & amplitudes $\Delta J\ \Delta H\ \Delta K_S$\\
17.43 & 16.35 & 15.82   & SAAO system\\
17.21 & 16.22 & 15.74   & corrected for $E_{B-V}=0.27$\\ \hline
\multicolumn{4}{l}{LDBK398: neighbour to \Sym}\\
\hline
17.27 & 16.25 & 15.63   & $\pm 0.02$ mag; mean \\
17.38 & 16.25 & 15.64   & SAAO system\\
17.17 & 16.13 & 15.56   & corrected for $E_{B-V}=0.27$\\
\hline
 \end{tabular}
\end{table}

Note that the $JHK_S$ mags given in the first line of Table~\ref{t:ir}
differ significantly from the values published in the 2MASS catalogue,
$J=16.91$ mag, $H=15.78$ mag and $K_s=15.29$ mag (the 2MASS quality flags are UBU
for this measurement). While this must be partially a consequence of
variability it is more important that 2MASS does not resolve two stars,
$\sim$2 arcsec from each other at the location of the target, into separate
sources. Our IRSF data show the nearby star
($\alpha_{2000.0}$ = 19:44:53.76, $\delta_{2000.0}$ = $-$14:51:46.2)
to be of similar brightness and its magnitudes are listed in Table~\ref{t:ir}.
Interestingly, this nearby star was identified as a carbon star by \citet{Let02}
using filters sensitive to absorption in the TiO and CN
bands; it is star number 398 in their list and identified as LDBK398 in our
Table~\ref{t:ir}.

\subsection{Spectrum}
\label{txt:spec}

The spectrum of \Sym\ presented in Fig.~\ref{fig:SySt} is that of a symbiotic
star (see, e.g. the atlas in Munari \& Zwitter (2002)).
\Sym\ shows both strong emission 
lines of \ion{H}{i} and \ion{He}{ii} \citep[e.g.,][]{Catalog} and a broad
emission feature at $\lambda$6830 \AA. Only symbiotic stars are known to
show the latter feature, which is due to Raman scattering of the \ion{O}{iv}
$\lambda\lambda$1032,1038 resonance lines by neutral hydrogen \citep{S89}.
The absence of forbidden lines in the spectrum of \Sym\ must be due
to collisional de-excitation, suggesting very high electron densities for the
gas surrounding the binary system, up to about
$N_e = 10^9$cm$^{-3}$ \citep[e.g.,][]{MAS97}. The high densities
\citep{GM95,P96} are consistent with the S-type symbiotic star
classification assigned below. The hot component of the binary is most
probably a white dwarf, but it is difficult to be certain of this without 
ultraviolet data.

\begin{figure}
    \begin{center}
    \includegraphics[angle=-90,width=8.5cm,clip=]{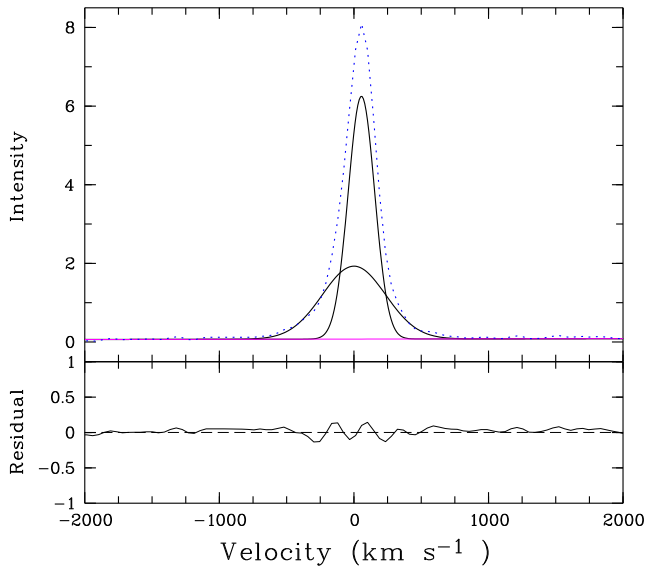}
    \caption{The H$\alpha$ profile of \Sym\ and the Gaussian two-component
    fitting to it. In the top panel the observed spectrum in the fitted
    region (dashed line), fitted Gaussians and assumed continuum level are
    shown.  The bottom panel illustrates the residuals after subtraction of
    the Gaussian fits and continuum level from the observed spectrum.  The
    narrow component has FWHM=5.1$\pm$0.1 \AA, which is close to the FWHM of
    the night-sky lines.  The wide component has FWHM=12.1$\pm$0.3
    \AA. The narrow and wide components are shifted relatively each other by
    1.2 \AA\ or 55 km s$^{-1}$.}
    \label{fig:Ha}
    \end{center}
\end{figure}

Balmer line ratios in many symbiotic nebulae indicate self-absorption
effects (because of high densities), making the application of standard
methods to estimate reddening difficult. Fortunately, we observed the
spectra of many \ion{H}{ii} regions (which are visible on Fig.~\ref{fig:FC}
as dark regions) around \Sym.  Using these data, we measured an average
$E(B-V)= 0.27\pm0.04$ for the region. Reddening estimates for \NGC\ 
range between $E(B-V)=0.20$ and 0.36 \citep[e.g.,][]{dV78,G96,Mas07,Gier06}
and it is obviously variable across the galaxy.
To estimate the temperature of the ionizing source (hot component) we used
the ratio \ion{He}{ii} 4686/H$\beta$ \citep{Ii82}, which gives us an upper
limit of $T^* \sim$130\,000~K since significant optical depth effects could
exist in the
Balmer lines. This calculated $T^*$ is close to typical values,
$T^*\sim100\,000$~K, for the hot components of symbiotic stars \citep{Murset91}.

The single-peaked H$\alpha$ profile, with a shoulder on the short
wavelength-side, is similar to those of well known symbiotic stars like AG
Draconis \citep{TT99,L_etal04}. The observed profile of the H$\alpha$ line
is shown in Fig.~\ref{fig:Ha}, where we also plot the result of Gaussian
two-component fitting. As can be seen from Fig.~\ref{fig:Ha} the H$\alpha$
line has extended low intensity wings which practically reach the continuum
level at about 750 km s$^{-1}$ (1000 km s$^{-1}$ in the case of AG
Draconis). \citet{TT99} suggested that the observed asymmetry of the
H$\alpha$ profile of AG Draconis is a sign of self-absorption and the
velocity in the wings should be close to that of the mass centre of the
system. In the case of \Sym\ the heliocentric radial velocity of the wings
is $-40\pm10$~km s$^{-1}$ and identical to that of the other Balmer lines
within the range of error. Comparison of \Sym's position and its optical
radial velocity with the 2D \ion{H}{i} velocity distribution (Figs.~2 and 3
from Weldrake, de Block \& Walter (2003)) shows agreement to within the
uncertainties of our velocity measurements.

\subsection{Classification}
\label{txt:class}

We can compare the IR colours with those of Galactic symbiotics in
\citet{WM92}, Figs.~2 and 6, which are on the SAAO system. There is no
sign of the extremely dusty, D-type, colours that are associated with
symbiotic Miras and we therefore classify this system as an S-type
symbiotic. On the other hand the colours are somewhat redder than most of
the Galactic S-types, particularly in $(H-K)$. This can be understood if the
red-giant in \Sym\ is a carbon star.  Indeed, the colours are not very
different from those of its neighbour which has been classified as a C star
(see section~\ref{txt:ast_phot}). Most of the Galactic systems have O-rich
giants.

\citet{Ita04} discussed $JHK$ photometry for red variables in the LMC
obtained with the IRSF.
Assuming that the difference in distance moduli between the LMC
and \NGC\ is 4.81 mag, then $K_0$ = 10.93 mag at the LMC distance.
This is in fact reasonably close to the value of $K_0$(LMC)=10.78 mag,
determined for the LMC from the mid-range of stars with P=142 days
and $(J-K)>1.4$ on sequence $\rm C'$ (according to Ita et al. Table~3
and assuming $A_K=0.02$ for the LMC).
This would suggest that the red giant in \Sym\ is pulsating in the
first overtone. It is about 1.2 mag brighter than the tip of the giant
branch (which is at $K=12.1$ mag in the LMC) and therefore clearly on the
asymptotic giant branch (AGB).

The colours of this star are very similar to those of C-rich variables in
the Galaxy as discussed by \citet{White06}. Although it has a somewhat
shorter period and is bluer than most of the stars from \citet{White06}, it
does fall very close to the $(J-H)_0/(H-K)_0$ relation illustrated in Fig.~8
of that paper. NGC\,6822 is known to contain a large number of carbon stars
(Letarte et al. 2002; Kang et al. 2006) and indeed, given its low
metallicity, we would expect most of its upper AGB stars to be C-rich. This
is because stars become C-rich when a sufficiently large number of carbon
atoms have been dredged-up to outnumber the oxygen atoms. Stars with lower
metallicity (or more strictly lower abundances of $\alpha$-elements) have
less oxygen to start with and therefore become C-rich soon after third
dredge-up starts.  Using the $(J-K)$ dependent bolometric correction from
\citet{White06} we find that
$m_{bol}=18.9$ (if the star were O-rich it would not be significantly
different) or $M_{bol}=-4.4$ mag and $M_K=-7.8$ mag. 

It is perhaps surprising that we do not see in the spectrum any stellar
features that might be attributed to the late-type giant, although this may
simply be a consequence of the low-level signal of the detected continuum. 
Carbon star SR variables typically have $5<(V-K)<8$, and possibly this one is
at the red end of this distribution as nothing is evident even at
$\lambda>6500 \AA$.  

\subsection{Symbiotic stars in the other galaxies}
\label{txt:Nearest}

\citet{Catalog} list the coordinates of the Magellanic Cloud
symbiotics, which have been discussed by \citet{MSV96} and
\citet{M04}, who noted that most of the cool components were AGB
stars, and by \citet{P07}, who discusses their 2MASS colours.  The LMC
contains three D-types (excluding Sanduleak's star whose status is 
unclear), two of which contain C stars, and four S-type, two of
which contain C stars. The 2MASS $K$-mags for the S-types are: 11.3, 11.5,
11.5 and 11.8 (the most luminous 2 are the C stars). The tip of the giant
branch (TRGB) in the LMC is at $K=12.1$ mag \citep[e.g.,][]{Ita04}, so, as
Miko{\l}ajewska pointed out, all the LMC symbiotics contain AGB stars.

In the SMC there are 6 S-type symbiotics, one of which contains a carbon
star. Their 2MASS $K$ mags are as follows: 10.8, 11.5, 11.5. 12.6 (C star),
12.6, 13.1. The SMC TRGB is around $K=12.7$ mag \citep[e.g.,][]{Ita04}, but
there is a significant line of sight spread, so most, possibly all, of these
symbiotics contain AGB stars.

The symbiotic star, Draco C-1, in the Draco dwarf spheroidal, shows emission
lines of \ion{H}{i}, \ion{He}{i} and \ion{He}{ii} that are very similar to
those in \Sym, but in this case carbon star absorption features are obvious
throughout the optical spectrum \citep{M91a}. The infrared colours
originally reported by \citet{M91b} ($J=11.60,\ H=11.35,\ K=11.40$ mag) are
much bluer than those of \Sym, and imply a significantly hotter star
with $M_K=-8.1$ mag. However, \citet{CH05} ($J=14.76\pm0.06\
K=13.90\pm0.05$ mag) and 2MASS ($JHK$, 14.38, 13.71, 13.46, mag) have very
different colours which would imply $M_K \sim -6.0$ mag. The TRGB in Draco is
around $K\sim 14.0$ mag \citep{CH05}, so these values still place
the giant on the AGB. It is not clear why Munari's original IR photometry is
so different from the 2MASS measurements.

\citet{IC10_SySt} have recently discovered a symbiotic star
IC\,10 SySt-1, whose spectrum is similar to that of He2-147. The latter is a
D-type symbiotic, but infrared observations are essential to distinguish the
new star's D- or S-type nature with certainty. The spectrum of the giant is
oxygen rich with a spectral type of M8III. If it were identical to He2-147,
with a pulsation period of 373 days \citep{SG07}, it
would have $M_K=-7.82$ mag \citep{WFvL08}. Assuming IC\,10 has a
distance modulus of $(m-M)_0=24.4$ \citep{IC10,Sa08} and a reddening
of $E_{B-V}=0.78$ \citep{Sa08}
the apparent mean $K$ mag of the symbiotic would be $\bar{K}\sim 16.8$.
With 2MASS data we have estimated the K is even brighter
($\sim 16.0$) and
conclude that \ICSym\ also contains giant companion in AGB phase.

\citet{WM92} pointed out that the $JHK$ colours of the S-type
symbiotics were unlike those of solar neighbourhood giants (their Fig.~6)
and speculated that they may be more like the metal-rich M-giants found in
the Galactic bulge. However, \citet{MGH05} found that abundance analyses did
not support this suggestion. The colours of AGB stars above the TRGB are
slightly redder than those of normal giants (this can be seen for stars in
the Fornax dwarf spheroidal in Fig.~3 (in the on-line version only) of
\citet{White09}, where the black points are AGB stars), and this is probably
the explanation for the colours of the symbiotic stars.

\section{Conclusions}

The combination of high excitation emission, in particular He{\sc ii} and
the Raman scattered feature at $\lambda$6830 \AA, together with the strong
IR flux unambiguously identifies \Sym\ as a symbiotic star, the first to be
found in NGC\,6822. The infrared colours further identify it as an S-type
symbiotic, with a carbon-rich AGB star pulsating in the first overtone with
a period of 142 days.

Evidence is presented that most extra-galactic symbiotic stars contain AGB
stars with luminosities greater than that of the TRGB. D-type symbiotics
contain Mira variables which are close to the top of their AGBs. Most
extragalactic S-types have luminosities above that of the TRGB and are
therefore on the AGB. The fact that one or two have luminosities close to or
below the TRGB does not prevent them from being early AGB stars. 
While this high fraction of AGB stars may be partly a selection effect it
seems very likely that many more of the Galactic symbiotic stars contain AGB
stars than has been appreciated to date.  

Our suggestion that
most, if not all, Galactic and extra-galactic symbiotics have AGB stars as
mass donors fits well with the view that mass is transfered through the
stellar winds, rather than via Roche lobe overflow, since the winds from AGB
stars are stronger than from normal giants.

\section*{Acknowledgments}

AYK is grateful to J.~Miko{\l}ajewska for discussions and valuable consultations
on symbiotic stars.
Some of the observations reported in this paper were obtained with the Southern
African Large Telescope (SALT), a consortium consisting of
the National Research Foundation of South Africa,
Nicholas Copernicus Astronomical Center of the Polish Academy of Sciences,
Hobby Eberly Telescope Founding Institutions, Rutgers University,
Georg-August-Università G\"ottingen, University of Wisconsin-Madison, Carnegie
Mellon University, University of Canterbury, United Kingdom SALT Consortium,
University of North Carolina -- Chapel Hill, Dartmouth College,
American Museum of Natural History and the Inter-University Centre for
Astronomy and Astrophysics, India.
We are grateful for the support of numerous people during the
SALT  PV phase.
This research has made use of the
NASA/IPAC Extragalactic Database (NED) which is operated by the Jet
Propulsion Laboratory, California Institute of Technology, under contract
with the National Aeronautics and Space Administration.


\bsp

\label{lastpage}

\end{document}